\documentclass[journal,final]{IEEEtran}
\usepackage{amssymb}
\usepackage{amsmath}
\usepackage{graphicx}
\usepackage{bm}
\usepackage{amsfonts}
\usepackage{cite}

\def\be{\begin{equation}} 
\def\ee{\end{equation}}

\begin{document}

\title{Challenges and Opportunities of Near-Term Quantum Computing Systems}

\author{A. D. C\'orcoles,~\IEEEmembership{Non-Member,~IEEE,}
A. Kandala,~\IEEEmembership{Non-Member,~IEEE,}
A. Javadi-Abhari~\IEEEmembership{Non-Member,~IEEE,}
D. T. McClure,~\IEEEmembership{Non-Member,~IEEE,}
A. W. Cross,~\IEEEmembership{Member,~IEEE,}
K. Temme~\IEEEmembership{Non-Member,~IEEE,}
\mbox{P. D. Nation}~\IEEEmembership{Non-Member,~IEEE,}
M. Steffen,~\IEEEmembership{Senior Member,~IEEE,}
J. M. Gambetta,~\IEEEmembership{Senior Member,~IEEE}
\thanks{All authors are with International Business Machines, IBM TJ Watson Research Center, Yorktown Heights,
NY, 10598 USA e-mail: (msteffe@us.ibm.com).}% <-this % stops a space
\thanks{Manuscript received August xx, 2019; revised yyyy zz, 2019.}}

\maketitle

\begin{abstract}
The concept of quantum computing has inspired a whole new generation of scientists, including physicists, engineers, and computer scientists, to fundamentally change the landscape of information technology. With experimental demonstrations stretching back more than two decades, the quantum computing community has achieved a major milestone over the past few years: the ability to build systems that are stretching the limits of what can be classically simulated, and which enable cloud-based research for a wide range of scientists, thus increasing the pool of talent exploring early quantum systems. While such noisy near-term quantum computing systems fall far short of the requirements for fault-tolerant systems, they provide unique testbeds for exploring the opportunities for quantum applications. Here we highlight the facets associated with these systems, including quantum software, cloud access, benchmarking quantum systems, error correction and mitigation  in such systems, and understanding the complexity of quantum circuits and how early quantum applications can run on near term quantum computers.
\end{abstract}

%\pacs{}

\IEEEpeerreviewmaketitle

\section{Introduction} 
\IEEEPARstart{Q}{uantum} computers can potentially solve problems that are considered intractable on even the fastest classical computers \cite{Nielsen00,Shor94,Grover97,Shor97,Yung14,QZoo}. They use a fundamentally {\em different} paradigm for performing calculations and solving problems compared with standard classical computers. The speed-up is achieved using quantum physics to explore correlations in problems such that the correct answer emerges at the end of a computation through constructive interference. This is obviously different from our standard picture of computation wherein each fundamental information unit is definitely in either the state 0 or 1 (we refer to references \cite{Gambetta17,Devoret13,Ladd10} as examples for a greater in-depth summary of quantum computing).

There is, however, a catch. The internal states of a quantum computer are fragile and susceptible to noise, introducing errors that lead to incorrect answers. Given the complexity and number of operations that are required for many typical quantum algorithms, it is believed that large-scale practical quantum computing has to incorporate at least some form of quantum error correction (QEC) \cite{Shor95qec,Nielsen00}. In the case of fault-tolerant quantum computing \cite{Aharonov97,Shor96ft,Preskill97,Kitaev03}, many quantum codes and techniques have been invented \cite{Campbell17}. Because simulating the full dynamics of quantum computers quickly becomes intractable as more qubits are added, QEC codes have been studied assuming simplified noise models such as Pauli noise. These simulations, together with assumptions about what is experimentally feasible, provide estimates of what would ultimately be required to operate various quantum algorithms using fully fault-tolerant computation \cite{Fowler12,Gorman17,litinski19,Gidney19}. Millions of qubits with relatively low physical error rates are predicted to be necessary to solve difficult problems. We will not expand further on fully fault-tolerant quantum computing for the remainder of this article but instead refer the reader to references \cite{Terhal15,Campbell17,Gambetta17}.

This article will focus on quantum computing with devices that are {\em currently} available or expected to be available in the near future. Various devices comprising 5-79 qubits have been made available to the public or exist as prototypes in laboratories \cite{ibm20q,google72q,rigetti,ionq}. Such devices have been referred to as ``noisy intermediate-scale quantum" (NISQ) \cite{Preskill18} systems, i.e., non-fault-tolerant devices comprising tens or hundreds of qubits. Such devices can be classified into two categories: a) devices constructed for a single demonstration experiment run by the team that created the device, and b) devices built to serve as general-purpose quantum systems for use by others. Designing a system requires consideration of many factors not relevant for a one-off demonstration. Five such factors, outlined below, are covered in detail in Sections~\ref{sec:eco}-\ref{sec:app}.

A system first needs to be designed to accommodate its intended users, enabling the functionality that they require to do their research and providing adaptability as their needs evolve. The breadth of quantum system users (physicists, computer scientists, engineers, chemists, developers, and others) requires multiple cloud systems and access interfaces for interacting with different systems at varying levels of abstraction. Examples of these include access levels for pulse and gate control, and ultimately for applications and systems comprising different connectivities. 

Second, while a demonstration can rely on special-purpose code, a system needs a complete software developer kit (SDK) providing a set of tools that can be used to develop novel experiments and applications. It is to this end that we have developed with the community Qiskit~\cite{Qiskit}, which consists of four fundamental elements: Terra~\cite{qiskit-terra}, Aer~\cite{qiskit-aer}, Ignis~\cite{qiskit-ignis}, and Aqua~\cite{qiskit-aqua}; each bring a specific set of features to the user. Terra provides the foundation for composing quantum programs at the level of circuits and pulses, optimizing them for the constraints of a particular device, and managing the execution of batches of experiments on remote-access devices. Aer gives access to high-performance quantum simulators to help us understand the limits of classical processors by demonstrating to what extent they can mimic quantum computers. Ignis offers a set of tools to better characterize errors, improve gates, and compute in the presence of noise. Finally, Aqua is where quantum algorithms are built and ultimately used in the context of applications. It provides translators to map problems from domains such as chemistry, optimization, finance, and artificial intelligence (AI) onto programs solvable with a quantum computer.

Third, it is important to establish a roadmap for the systems. Much like roadmaps for classical systems, a roadmap for quantum systems provides a community-chosen benchmark to facilitate comparisons across systems and demonstrate progress over time. For quantum computers, many individual metrics are commonly accepted as ingredients for a better quantum system but, as of today, a single community-wide accepted benchmark does not exist. Quantum Volume (QV) has been proposed as a potential benchmark that incorporates many of the individual metrics (number of qubits, connectivity, gate set performance, and compiler and software stack performance) into a single hardware-agnostic metric~\cite{Cross2018}. We have shown improved QV over the past two years, and strive toward continued improvements.

Fourth, to extend the computational reach of these shallow depth quantum circuits, error mitigation has been proposed as a technique to increase the accuracy of measured observables~\cite{Temme17,Li17}. Error mitigation is a term used to express methods by which the impact of error can be reduced (or mitigated) without requiring full fault-tolerant quantum codes. This approach to reducing errors in the absence of full fault tolerance is still very much exploratory, but we view it as an important component of near-term quantum systems.

Fifth, quantum computers without fault tolerance will likely be limited to implementing algorithms with short-depth quantum circuits~\cite{Bravyi2018}. In these algorithms, a series of quantum gates are applied and then the qubits are measured. The outcomes of these circuits are used to compute an observable or sample from a probability distribution of interest. This limited model is believed to be computationally hard for classical machines~\cite{Terhal2004}, and recently it has been shown to have an unconditional separation between classical and quantum computers~\cite{Bravyi2018}. Using this model, researchers have explored applications in quantum machine learning~\cite{Havlicek18} and quantum chemistry~\cite{Kandala18}. 

In addition to these five system aspects, the quantum hardware itself is, of course, a key component of a quantum system. A discussion of specific challenges associated with our quantum hardware of choice (superconducting qubits) is beyond the scope of this article. On this subject we instead refer the reader to additional resources such as Refs.~\cite{Gambetta17,Devoret13,Krantz19,Wendin17} for superconducting qubits, and references within \cite{Gambetta17} for other qubit implementations.

Nonetheless, we stress the importance of engineering systems capable of operating over prolonged periods of time. Rudimentary experiments are notoriously unstable and often barely capable of gathering sufficient data for a scientific publication. Both external influences and internal device noise can cause parameters to quickly drift on experimental timescales, potentially rendering a device unsuitable for use in a quantum system due to the prohibitive amount of device calibrations required. 

From these five aspects, it is evident that developing a complete, user-friendly, cloud-accessible quantum system necessitates considering a rich landscape of design aspects. Successfully implementing all of the ingredients simultaneously to achieve this goal is no small task. This article reviews these considerations in the next five sections. For quantum applications, particular emphasis is placed on describing algorithms for quantum machine learning and quantum chemistry, because they are examples of applications that can be mapped to  short-depth circuits that are believed to be hard for a classical computer to simulate and are currently areas of great interest. 

%Most importantly, we must foster an ecosystem to create a user base. Cloud-accessible quantum systems enable users to perform those experiments that they consider important for their research, whether that is research performed at industrial partners or at universities. As discussed, this requires the underlying hardware and software to exhibit the necessary flexibility, and crucially stability, to enable desired experiments on near term quantum systems. 
\section{Cloud quantum systems and user access levels}
\label{sec:eco}

Although experimental research in quantum computing has been active for over two decades, it was not until the mid-2010's that it became possible to physically connect a handful of superconducting qubits together to implement small multi-qubit tests with sufficient fidelity for meaningful results. In 2016, IBM built a quantum processor composed of five superconducting qubits and integrated it into a system called the IBM~Quantum Experience\cite{QX16} available for use via cloud access. Almost immediately after launch, research papers were published based on results obtained through this cloud access to a quantum device. It demonstrated one key aspect that we believe is important for the future as well: there is already demand from physicists, scientists, developers, and many others to access and test various aspects of quantum computing, even if the systems are still small, comprise only a few qubits, and suffer from noise levels worse than the fault-tolerance threshold.

Since the initial release of the 5-qubit backend, the IBM~Quantum Experience has hosted thirdteen unique quantum systems made available as backend services either to the public or to members of the IBM~Q Network \cite{ibmqnetwork}. Over time, user executions have increased to 28 million, culminating in over 180 research papers exploring areas in quantum information science. The proliferation and quality of enabled research serves as an affirmation of the community-wide demand for a variety of quantum systems.

\begin{figure}[tbp!]
		 \centering
	\includegraphics[width=0.5\textwidth]{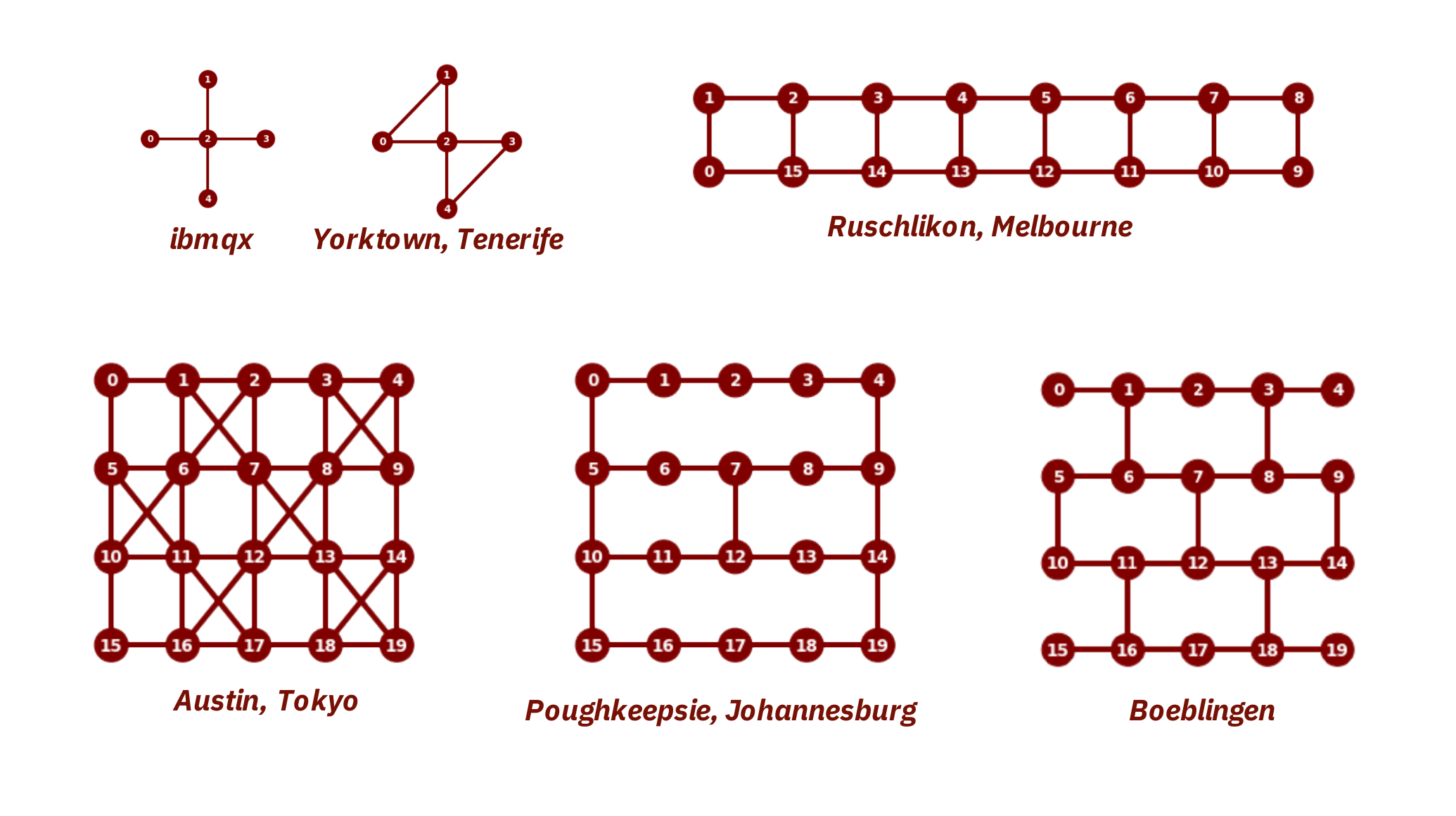}
		 \caption{Examples of several IBM cloud accessible devices. The top left 5-qubit device was the first one made available via the IBM~Quantum Experience \cite{QX16}. The one to the right of it was made available after including additional entangling gates between two pairs of qubits. A 16-qubit device was made available approximately a year after the first device. The devices in the bottom row show three variations of 20-qubit devices available to members of the IBM~Q Network \cite{ibmqnetwork}.}
		 \label{fig:device}
\end{figure}

\begin{figure}[bp!]
		 \centering
	\includegraphics[width=0.5\textwidth]{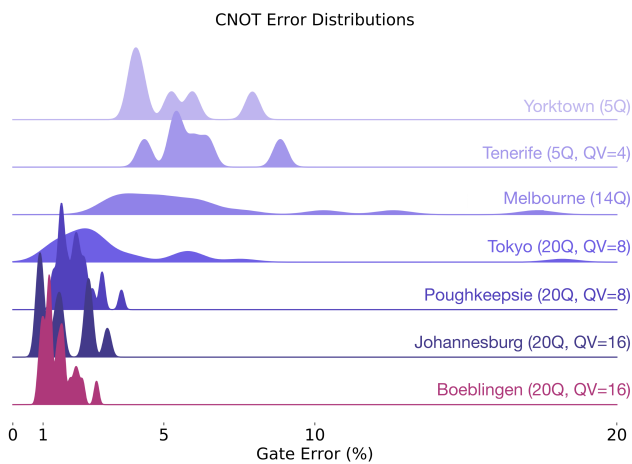}
		 \caption{Controlled-NOT (CNOT) gate error distributions for a variety of IBM devices. Beginning with the earlier devices (top rows), the average error rates remained quite large but have improved with continuing research. The bottom row represents the device shown in Fig.~\ref{fig:20qdevice}. The error reductions are the result of improved gate fidelities and increasing coherence times \cite{Sheldon16,Gambetta17a}, as well as a better understanding of spectator qubit errors.}
		 \label{fig:errorfig}
\end{figure}

The qubit connectivity maps and the distribution of two-qubit error rates for a few of the available IBM backends are shown in Fig.~\ref{fig:device} and Fig.~\ref{fig:errorfig}, respectively. The variety of devices allows us to explore user preferences and device performance for different connectivities. 

In these near-term systems, imperfections and connectivity can have a large impact on the performance of different algorithms. More connectivity allows users to explore circuits that entangle the qubits in fewer steps, but often at the price of hurting gate fidelities or inducing spectator errors \cite{McKay19, Takita16}, i.e., errors that can occur on qubits that are still passively connected but otherwise not directly involved in a particular quantum operation. As we progress through this period of near-term quantum systems, we must evolve towards co-design of the quantum circuits users want to implement, as well as the connectivity that is physically built into the systems. 

\begin{figure}[tbp!]
		 \centering
	\includegraphics[angle=-90, width=0.35\textwidth]{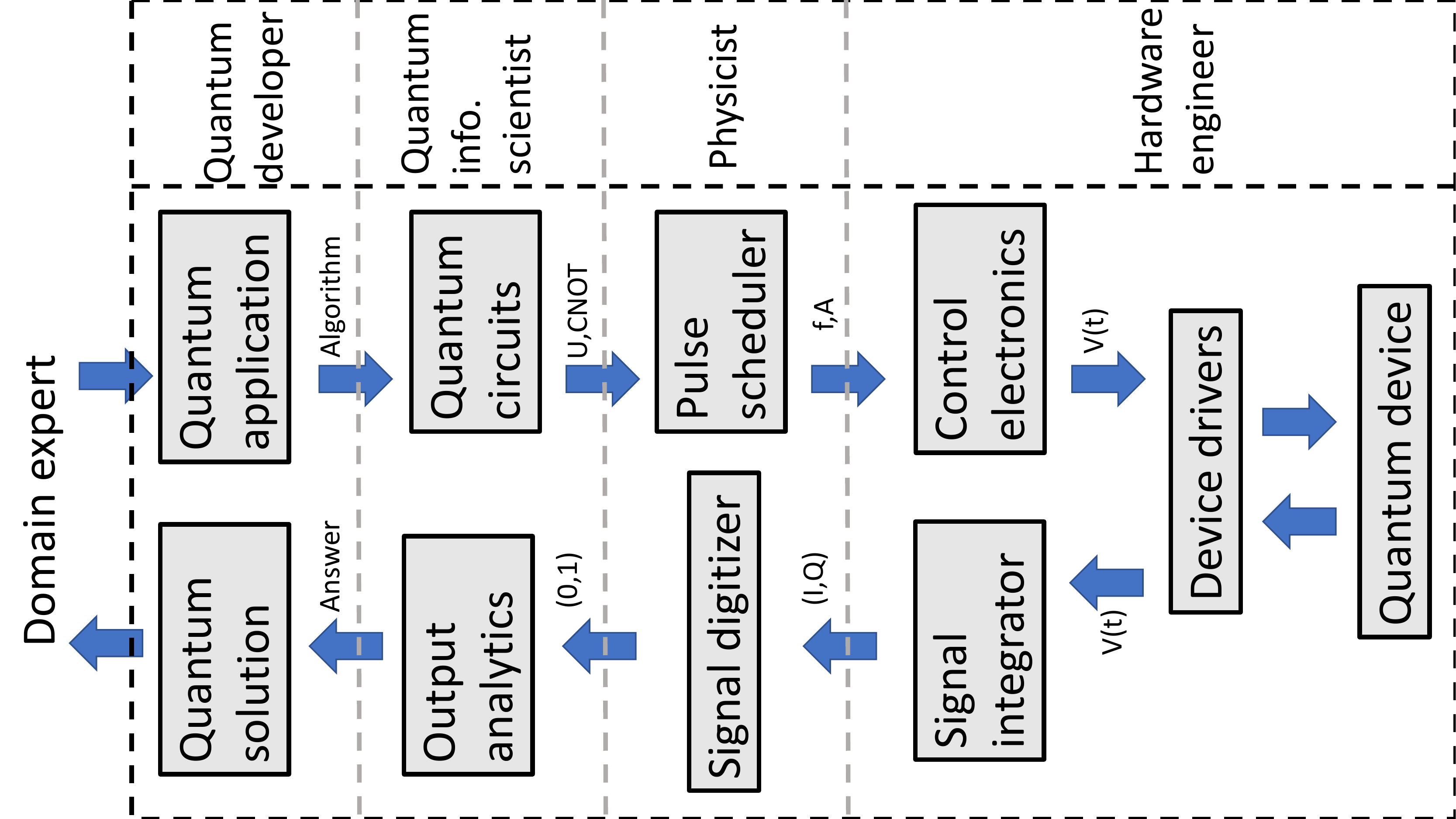}
		 \caption{A schematic overview of a cloud-enabled quantum computer and user access levels. At the high end of the stack, quantum developers create quantum applications executed in the form of algorithms that are translated into quantum circuits at the next level. Quantum information scientists have access to implement specific quantum circuits of interest, which are translated into sequences of quantum operations at the next lower level. Here we envision that physicists can implement specific pulse-level experiments, such as optimal control or gate-level research. These instructions are sent to the quantum hardware via a set of control electronics in the form of frequencies and drive signal amplitudes. The quantum hardware is accessible by hardware engineers. At the conclusion of an experiment, the hardware passes the readout signal in the form of readout voltages to a signal integrator. The integrator signal is represented as the I and Q quadratures of the readout signal, which get digitized at the physicist access level to a logical 0 or 1. The logical bit stream is passed to output analytics for the quantum information scientists to analyze results of the implemented quantum circuits. Finally, the answer of a full quantum application is sent back to the quantum developer. A domain expert is the expected end-user of a fully developed stack. This representative stack demonstrates how different levels of access are seamlessly possible.}
		 \label{fig:access}
\end{figure}

Furthermore, user feedback showed clear interest in more than one level of access to a quantum device. We consider three definable fundamental user classes for various levels of cloud access (see Fig.~\ref{fig:access}): the quantum physicist, the quantum information scientist, and the quantum developer.

The quantum physicist possesses a deep understanding of the underlying device physics, and would like to explore more practical technical details, such as optimal control techniques, novel pulse-shaping approaches, techniques to quantify the underlying system Hamiltonian, and error mitigation methods. These users want more of the nitty-gritty details, and the ability to examine device-level properties of the system, e.g., control over the frequency, timing, pulse shapes, and measurement integration kernels that are sent to the experiment. To properly meet this level of user access we have defined the \textit{OpenPulse}~\cite{McKay18} framework, along with a corresponding set of tools in Qiskit described later in this review. Succinctly, \textit{OpenPulse} provides the bare metal access level for users.

The quantum information scientist has a deep understanding of quantum circuits and wants to explore how these circuits run on near-term devices. A circuit can be implemented multiple ways, in terms of the fundamental gates. Finding the optimal solution is a computationally difficult task, and is therefore an important research topic. These users are also interested in exploring error-correction primitives, such as parity checks and conditional operations that depend on these multi-qubit measurements to investigate how entropy is taken from the system. For this level we have defined \textit{OpenQASM} \cite{Cross17} and the corresponding tools in Qiskit.

The third user class, the quantum developer, wants to see how quantum applications work on quantum computers.  They want to run circuits based on an application and receive the outcome as quickly as possible. They are not necessarily interested in how the circuit is implemented; they are focused on the results of the quantum computation, and how it can be used in an application of interest. 

\begin{figure}[bp!]
		 \centering
	\includegraphics[width=0.5\textwidth]{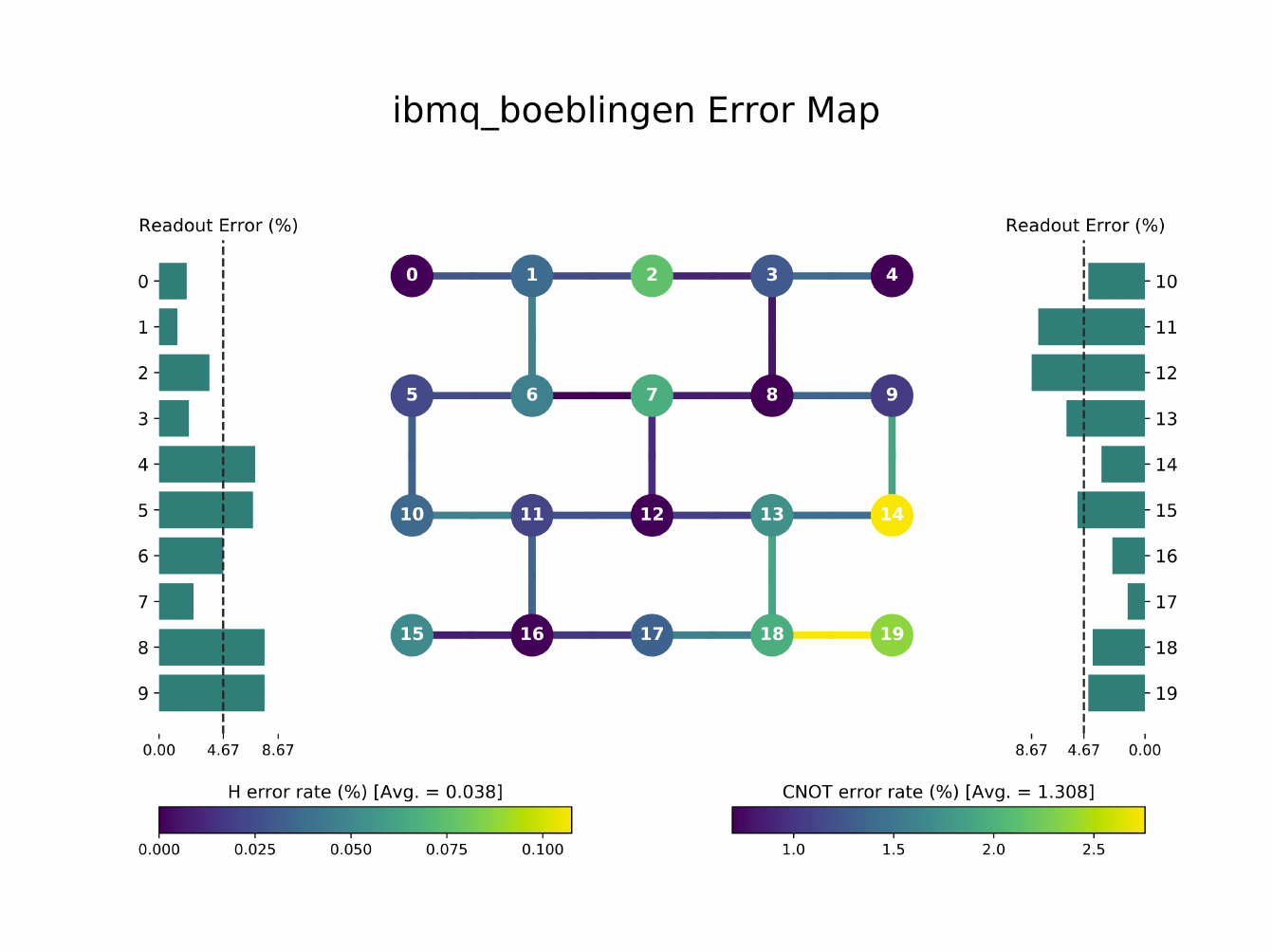}
		 \caption{Error map for the Boeblingen 20-qubit device at IBM. Qubits are represented by dots and connected to other qubits via lines. The colors of the qubits represent the measured Hadamard error rate. Z-rotations are implemented in software \cite{McKay17}. For this device, an average Hadamard error of $0.038 \%$ is measured. The color of the lines represents the measured two-qubit CNOT error rate between given pairs of qubits. For this device, an average CNOT error of approximately $1.3 \%$ is measured. The vertical bars to the left and right of the device plot the readout error rate, with an average readout error of approximately $4.7 \%$.}
		 \label{fig:20qdevice}
\end{figure}

In parallel, it is also important to provide to each of these users the data appropriate for their respective level. The physicist needs access to device-level specifications, while the quantum information scientist needs the error rates for the calibrated quantum gates and operations. Device specifications are the fundamental properties of the device (e.g.,
coherence times, qubit frequencies, and crosstalk), while error rates depend on the pulses that represent the gates, and include metrics such as average single-qubit errors, two-qubit gate errors, spectator errors, assignment (or readout) errors, and readout crosstalk errors. Cloud-enabled devices typically list most (if not all) of these metrics; see the example in Fig.~\ref{fig:20qdevice} for the backend properties needed by a user of the quantum information scientist persona. By tracking the metrics, we can gauge the importance of any particular metric (or combination thereof), in order to improve the overall quality of experiments.

\section{Qiskit and compilation}
\label{sec:qk}

An unprecedented acceleration of research and development in quantum computing has occurred in recent years, chiefly enabled by wide public access to cloud quantum computers. The software stack plays a key role in taking advantage of these systems and enabling quantum information science as a whole~\cite{Qiskit,pyquil,cirq,projectq,scaffcc}. In this section we discuss Qiskit, a software suite for near-term quantum computing.

We will pay special attention to the compiler as an indispensable part of any quantum computing system. Our description is focused on compilation strategies tailored to near-term noisy systems. Compiling for fault-tolerant machines is a vast area of research in itself, and we refer the interested reader to references~\cite{dawson05,ross2014,amy2014,bocharov2015,javadi2017} for further reading.

\subsection{Qiskit architecture}
\begin{figure*}[tbp!]
		 \centering
	\includegraphics[width=\textwidth]{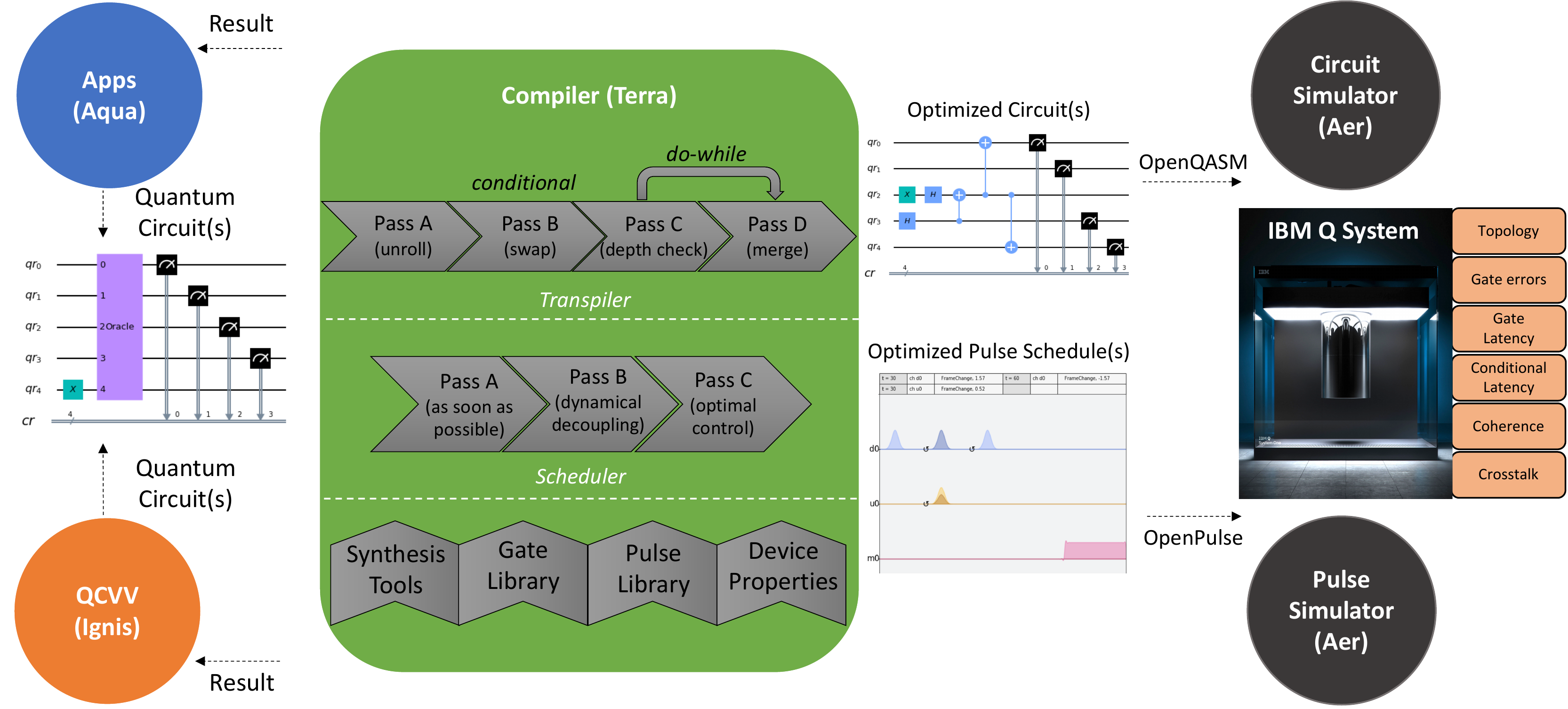}
		 \caption{Architecture of Qiskit. Aqua and Ignis produce circuits for different tasks (algorithms and applications, or device characterization and validation (QCVV), respectively). The IBM~Q systems and Aer simulators are backends that execute quantum circuits or pulse schedules. The Terra compiler is the bridge that translates and optimizes for a given backend, and is comprised of modular pass-based circuit optimizers (Transpiler) and pulse optimizers (Scheduler). Some example passes are shown. Efficient high-level synthesis methods, access to a library of pre-computed gate and pulse equivalents, and information about device constraints and properties all increase compilation quality.
		 \label{fig:qiskit}}
\end{figure*}

Figure~\ref{fig:qiskit} shows the overall architecture of Qiskit.
As a software development kit and research tool, Qiskit is comprised of \emph{elements} that we deem important in the journey toward quantum advantage. For the sake of completeness, we review them again. The first element, \emph{Terra}~\cite{qiskit-terra}, provides the foundations and language to describe quantum computations at different abstraction levels (circuits or pulse schedules), and to compile and optimize them for specific machines. The second element, \emph{Aer}~\cite{qiskit-aer}, provides scalable and realistic simulations of quantum systems, and is invaluable to understanding the complexity of different computations, as well as how they behave under certain noise assumptions. The third element, \emph{Ignis}~\cite{qiskit-ignis}, provides tools to characterize quantum devices and mitigate the effects of noise on them. The fourth element, \emph{Aqua}~\cite{qiskit-aqua}, is a library of quantum algorithms and translators from near-term application domains (such as chemistry and AI) to quantum circuits.

Since quantum software is so new, there are many unknowns in how the software stack should be configured for each particular setting. In addition, active research is in progress in all aforementioned areas. For this reason, Qiskit has a highly modular architecture, easily extensible at all levels. This includes adding new circuit optimization passes, new noise models for simulation, new algorithms, and new noise characterization and mitigation methods.

The Qiskit compiler is primarily composed of two parts, the \emph{transpiler} and the \emph{scheduler}. The transpiler is a circuit-rewriting toolchain, designed to optimize circuits, both in the abstract and for particular backends. The scheduler converts circuits written for a given device into the sequence of pulses executed on that device.

In the transpiler, multiple circuit analysis and transformation ``passes'' can be strung together to yield a custom circuit optimization pipeline. A typical sequence might include unrolling the circuit gates to a particular native gate set, allocating ancilla qubits, swapping qubits so that entangling interactions match the device topology, merging consecutive gates into simpler ones, analyzing commutation relations and cancelling non-adjacent gates, analyzing the circuit depth, and repeating a couple of optimization passes until the circuit depth reaches a fixed point. A ``pass manager'' sequences the user's desired passes, keeps internal context about the progress, and ensures that the control flow of passes is implemented correctly. 
Similarly, the scheduler contains passes that convert a circuit to a sequence of pulses with specific timing --- for example, using an as-soon-as-possible or as-late-as-possible scheduling method --- and can optimize them further using methods such as dynamical decoupling~\cite{cywinski2008,viola1999} or optimal control~\cite{Khaneja05}.

\subsection{Compiling for near-term machines}
Near-term quantum hardware is severely limited in what it can compute. Errors can build up rapidly during the execution of a program, and can render a computation useless. In contrast to classical compilers, for which the goal is to transform a program to run faster, the primary goal of a compiler for near-term quantum computers is to combat these errors. Therefore, a good quantum compiler must ensure that an input program is translated into the most efficient equivalent of itself, squeezing the most out of the available hardware.

Some steps in the compilation process are necessary to run the program in the first place. For example, high-level program routines, such as an abstract unitary evolution, must first be \emph{synthesized} into a quantum circuit~\cite{synthesis1,synthesis2,synthesis3,synthesis4}. A circuit must be transformed to conform to the hard constraints of a device, such as which qubits can interact with one another, or which gates are natively supported~\cite{mapping1,mapping2,mapping3,mapping4,mapping5}. Finally, circuits must be translated into pulses that control the qubits~\cite{Sheldon16,McKay17}.

Beyond this, an {\em optimizing compiler} should focus on the soft constraints given by the physics of the device, and optimize within that space. For near-term quantum computers, seemingly small optimizations such as reducing the two-qubit entangling gate (e.g., CNOT) count by 15\% can yield dramatic improvements in the final fidelity of computation. The compilation problem in general is NP-hard~\cite{maslov2008}. Finding optimal layouts of program qubits on the device, or finding optimal swapping routes between the hardware qubits, can be done by solving subgraph isomorphism and token swapping problems, respectively. We may be able to find optimal solutions for small systems, but soon we need to devise effective heuristics.

An optimizing compiler must generally be aware of the set of constraints and parameters within which it is trying to optimize. To first order, these can be generic truths, such as the fact that two-qubit gates have higher errors than single-qubit gates, or that qubits lose their information if the program length (i.e., quantum circuit depth) is too long. Given these constraints, general optimization {\em objectives} are defined for a quantum compiler, such as minimizing the circuit depth or the number of entangling gates. This has been the traditional approach to circuit optimization for more than a decade~\cite{optimization1,optimization2,optimization3,optimization4}.

While effective, circuit depth and gate count are only pseudo-objectives to simplify reasoning about the quality of a compiler's optimizations. In reality, what matters is the fidelity of a computation when run on actual quantum hardware. Every quantum device is different, and thus benchmarking and characterizing the system are critical to successful compilation. As an example, it is often taken for granted that lower circuit depth is better. This has resulted in trying to parallelize gates as much as possible~\cite{scheduling1,scheduling2,scheduling3}, which may yield bad results on a high-crosstalk system. Conversely, randomized compiling~\cite{wallman2016} prolongs circuit depth by inserting extra gates, yet the effect of these gates is to randomize and mitigate coherent errors, achieving a better overall fidelity.

The key takeaway is that compilers for noisy quantum computers excel when more information is made available to them from the device. With cloud-access quantum computers, the field of quantum computer science is moving towards evaluating the effect of compilation strategies on real hardware, rather than based on objectives that may not be comprehensive. In the IBM~Q ecosystem, device properties are shared openly and can be benchmarked, resulting in a flurry of recent compiler innovations~\cite{tannu2018,murali2019noise,murali2019triq,nishio2019,finigan2018}. Pertinent hardware characteristics include, but are not limited to: qubit topology, native gate sets, gate error rates, latencies of gates, readouts and feed-forwards, qubit lifetimes (decoherence and relaxation), and crosstalk errors. 

A key observation in compiling for noisy quantum computers is that, since errors always exist, it may not always be worth performing a numerically exact compilation. Alternatively, {\em approximate compilation} aims to approximate a computation (a unitary) by some numerically close alternative, in order to potentially save significant resources \cite{Cross2018,peterson2019}. If the reduction in error due to the shorter alternative is more than the loss of precision in the approximation, then this trade-off is worthwhile. Evaluating this trade-off is again dependent on the exact characteristics of the device.

Finally, verification of the compiler becomes a serious challenge even in the near term, as verification of general circuit transformation on circuits of roughly 50 or more qubits is impractical. Consequently, expansive testing of smaller cases or formal verification methods~\cite{amy2017,shi2019} will be essential.

Quantum compilers have benefited from decades of classical compiler design, yet the new domain creates new challenges and opportunities. For example, commutation relationships among quantum gates provide additional flexibility, compared to a classical program for instructions to be reordered, merged, or cancelled~\cite{scheduling1,scheduling2,scheduling3,itoko2019}. In contrast to classical computing, in which a program can be compiled once and reused thereafter, quantum programs must often be recompiled, as device properties change over time. Lastly, traditional ISA boundaries may not work well on near-term devices, as they may sacrifice some efficiency in favor of abstraction~\cite{shi2019optimized}. Designing an industrial-scale compiler suitable for the coming generation of quantum computers remains an exciting and hard task.
\section{Benchmarking near-term devices}
\label{sec:benchmarking}

Benchmarking quantum systems will be a necessity to measure progress. Assuming several physical realizations of quantum computers will emerge over time, there must be a way to quantify their respective performance much like classical benchmarks. While benchmarking appears to be an obvious requirement, it is far less obvious how to devise a rigorous set of metrics applicable to quantum computers. 

There are a number of important factors for formulating an appropriate quantum benchmark:
\begin{itemize}
\item {\em Number of qubits}: More qubits are required to solve increasingly difficult problems, and thus, everything else being equal, the more qubits a quantum computer has, the more computational power it has. Systems with several tens of qubits can be simulated on a classical computer, and therefore having a few qubits will not be beneficial in the long run.
\item {\em Connectivity}: How qubits are connected to one another matters. At one extreme, qubits connected on a line would require a significant overhead for any randomly selected gate between any random pair of qubits. At the other extreme, if all qubits are connected to each other, there is no additional overhead for a randomly selected gate between any random qubit pair. However, at the hardware level connectivity matters a lot, and greatly influences metrics such as crosstalk, fidelity, etc. It is important to strike a balance between connectivity and overhead for a given application.
\item{\em Error rates}: Quantum operations that feature lower errors rates are generally better. A critical component associated with error rates are spectator errors (errors on qubits that are not participating in the applied quantum gate). The spectator errors can significantly degrade or even dominate overall circuit performance. For example, while a two-qubit gate might lead to very low errors on the two qubits involved, it is possible that another qubit might undergo significant errors due to the application of the two-qubit gate.
\item {\em Gate set}: The choice and performance of the underlying gate set is important.  A large set of gates reduces the overhead to synthesize arbitrary gates or move quantum information, but also requires far more complexity from a calibration and stability standpoint.
\item {\em Compilers and software stack performance}: Compilers are critical for optimal translation of circuits to the underlying hardware. A compiler needs to consider and optimize over the device connectivity, and potentially even variations of gate fidelities and spectator errors across the device.
\end{itemize}

IBM has devised a benchmark called {\em Quantum Volume} that balances all of the ingredients above \cite{Cross2018}. We believe this system-agnostic metric provides a way to compare devices across different physical implementations, and measures qualities, such as low error rates, that are ultimately necessary for a practical quantum computer.

The Quantum Volume measures the largest model circuits the quantum computer can successfully run. A model circuit consists of random two-qubit gates acting on random pairs of qubits, and has as many parallelized layers of these gates as it has qubits. The model circuits are compiled to the particular quantum system. A given run is considered successful if the observed measurement outcome is in the upper half of the ideal output probability distribution. To claim that the quantum volume exceeds some value for some system, we are required to succeed for more than two-thirds of the runs on a given number of qubits.

We wish to particularly highlight the role of circuit compilation in the Quantum Volume, because a full quantum system is the combination of the qubits, gates, control electronics, and the software stack that optimizes for those components. The Quantum Volume provides a way to benchmark the whole quantum computing system, including the optimizing components of the software stack.

Choices for universal benchmarks will evolve as the community continues to learn more about near-term devices \cite{blumekohout19,erhard19,hamilton19}, but the quantum volume is an accessible and measurable quantity that tracks progress on current devices. We have released an open-source library for measuring Quantum Volume in Qiskit. The largest Quantum Volume measured thus far is 16, measured first on the 20-qubit Johannesburg system \cite{Cross2018}, and more recently on Boeblingen, Fig.~\ref{fig:20qdevice}.

\section{Error mitigation and correction}
\label{error_mitigation}

For decades researchers have understood that decoherence would limit the duration of useful quantum computation \cite{Unruh95} and have devised many techniques for overcoming noise. Today we are keenly aware of the impact of decoherence and control error on the size and accuracy of quantum computations. The way forward is necessarily a mixture of approaches. Foremost, we must understand and reduce the fundamental error sources in our hardware, control systems, and environment. Beyond this, we must correct the remaining errors and/or mitigate their effect on the quantum computer's accuracy in a resource efficient way. These are among the central research challenges for the foreseeable future.

It is by now well-known that the principles of quantum error correction allow errors to be dramatically and efficiently suppressed in theory \cite{Shor95qec,Nielsen00,Aharonov97,Shor96ft,Preskill97,Kitaev03}, so that the computational time can extend well beyond the coherence time. This requires physical error rates to be low enough and noise to be sufficiently uncorrelated. When that happens, fault-tolerant gates can successfully limit the spread of errors and quantum error correction procedures can remove entropy faster than it accumulates. If error rates are only modestly below the threshold error rate, the additional space and time to implement fault-tolerant gates can be prohibitively large. Furthermore, topological codes \cite{Dennis02}, which are among the most well-suited to planar quantum computing architectures, are expected to correct errors very well, but protect qubits by encoding each of them into a large number of physical qubits.

In the near term, we are unlikely to have both sufficiently low error rates and sufficiently many qubits to implement a fault-tolerant quantum computer. Nevertheless, these near term systems present an early opportunity to research error mitigation and error correction in real noise environments. On the one hand, quantum error correction experiments spur development, confirm predictions, and expose facts about detecting realistic errors. On the other hand, error mitigation experiments have low overhead and significantly improve computational results today, so they are eminently practical. Error mitigation can improve estimates of expectation values, which can be important in explorations of quantum advantage -- for example, as eigenvalues of molecular Hamiltonians or Kernels in classification problems addressed by quantum machine learning algorithms. However, unlike quantum error correction which removes entropy, error mitigation cannot extend the computation far beyond the coherence time.

We now focus our attention on error mitigation schemes, which are more recent and less well known that error correction schemes. To date, there have been two general-purpose error mitigation schemes developed. The first, zero-noise extrapolation, was developed independently in the works of \cite{Temme17,Li17}, and second, probabilistic error cancellation was introduced in \cite{Temme17}. In zero-noise extrapolation, the output from a circuit of interest is re-measured under different amplified noise strengths. The measured expectation values from these noisy runs can then be recombined to extrapolate to an estimate of the expectation value at the zero-noise limit that is more accurate than the best individual run. With measurements at an increasing number of noise strengths, a Richardson extrapolation can be employed to increasingly suppress the noise contributions to the zero-noise estimate. Temme \textit{et al.}~\cite{Temme17} showed that such noise amplification could be achieved by stretching the time evolution of the quantum state, under the influence of the time-dependent drives that constitute the quantum circuit. Under the assumption of time-invariant noise, the stretch factor for the time evolution is equivalent to the noise amplification factor. Beyond this assumption, no further characterizations of the noise models are required, making this extremely attractive for experimental implementations. This method was demonstrated and integrated into a variational algorithm in the experiment of \cite{Kandala18}, using superconducting qubits and all-microwave gates. It was also employed to improve the performance of a binary classifier realized on the same device \cite{Havlicek18}. 

A second, general-purpose error mitigation scheme, also proposed in \cite{Temme17}, is termed probabilistic error cancellation or quasi-probability decomposition. In this method, every well-characterized noise channel in a quantum circuit is acted upon by its inverse. While implementing the inverse noise channel is in itself an unphysical task, it was shown that an ``average" error-mitigated estimate of the outcome can instead be obtained by sampling from an ensemble of noisy circuits with probabilities related to the coefficients of the inverse noise map. The variance of the error-mitigated estimate is related to the number of noisy circuits sampled and measured. In contrast to the zero-noise extrapolation technique, a key experimental challenge here lies in the characterization of noisy gates employed in the quantum circuit. For up to two-qubit experiments, this method was recently realized for superconducting qubit~\cite{song2018} and trapped ion architectures~\cite{zhang2019}, both employing gate set tomography for noise characterization. 

In addition to these techniques, other methods have been proposed that are more problem-specific. The quantum subspace expansion method~\cite{QSE,Colless2018} involves the measurement of additional excitation operators for variational ground states, and in addition to providing excited state energies, also mitigates on energy estimates. Other recent approaches to error mitigation for fermionic problems rely on the conservation of ``known quantities", such as particle number ~\cite{McArdle2019,Bonet2018,Delfterrormit}. Such symmetries can be enforced by using ancillary qubits to perform stabilizer checks.

Error mitigation is still in its infancy but has shown some promising first steps. As we look forward, we hope that access to near term systems will enable new error mitigation and correction techniques at the intersection of theory and practice. Ultimately, we believe that quantum error correction and fault tolerant design will still be necessary. Therefore, continued experiments such as demonstrations of Bell state parity measurements~\cite{Corcoles15}, stabilizer measurements~\cite{Takita16}, error detecting codes~\cite{Linke17,Takita17}, and other codes~\cite{Gong19} are critical for understanding how to protect encoded quantum information in the long term~\cite{chamberland19}. We anticipate the theory and practice of error correction and mitigation to continue to develop together in the future and that new ideas will emerge.
\section{Quantum applications on near-term quantum systems}
\label{sec:app}

The relevance of a quantum computer is derived from the algorithms that can be performed on it. For some problems, such as for example factoring integers \cite{Shor94} or simulating quantum mechanics \cite{zalka1998simulating}, quantum algorithms have theoretical guarantees to drastically outperform any known classical algorithm. It is important to state that not every problem that is challenging for a classical computer will benefit from a quantum speed up. This means that the applications for a quantum computer have to be identified individually and a specific quantum algorithm has to be developed for them. Up to this point, the set of algorithms that can be shown to outperform classical computers \cite{Algorithms-Zoo} all depend on an architecture that is fully fault tolerant. The quantum hardware that is currently available is not yet at a stage to run fault tolerant computations. Nevertheless, making current hardware available to the research community allows for the investigation of quantum algorithms that have the potential to run on near-term quantum devices. Due to device imperfections and decoherence, we expect that such algorithms will be comprised of quantum circuits that are of shallow depth. To tackle a complex computational task, some of the computation that does not benefit from a quantum speed up can be outsourced to a classical computer. Examples for such quantum - classical hybrid schemes are variational algorithms for quantum many-body systems \cite{peruzzo2014,Kandala17,Kandala18} and machine learning \cite{Havlicek18}. Such shallow depth variational hybrid- algorithms can be understood from the following picture; The classical computer tries to find the best quantum circuit, limited in size to a depth determined by the noise, to perform a particular computational task. The task could for example be the preparation of an approximation to the ground state of a Hamiltonian or the construction of a classifier in machine learning. This simple scheme has opened up a pathway to trying heuristic algorithms, that don't come with any performance guarantees on current quantum hardware. 

The development of classical algorithms has greatly profited from the wide availability of computational hardware. Many heuristic algorithms were found by trial and error and come without performance guarantees. It is therefore reasonable to likewise follow an experimental route in the search for applications that could benefit from a quantum computer. However, there is an important difference between the development of classical algorithms and quantum algorithms. 
Not all quantum circuits can lead to a quantum advantage. If the quantum algorithm can be efficiently simulated on classical hardware \cite{Tang19,GLT19,BG16,StabRank,AG04,Razborov04,BMS17}, it can not provide a computational advantage. The advantage of quantum computers is based on the complexity of the algorithm and not on the quantum computers ability to perform fast operations. It is therefore paramount to ensure that the quantum algorithm is based on a circuit that can not be efficiently simulated on a classical computer. 

To ensure classical hardness of simulation is of particular importance when performing algorithms on a small number of qubits that are subject to noise, since a quantum advantage may not be immediately apparent. The first fundamental question that arises is, whether, and under which circumstances, a shallow depth quantum circuit can provide a computational advantage. This question was recently addressed and answered by Bravyi \textit{et al.}~\cite{Bravyi2018}. In their work, Bravyi \textit{et al.} demonstrate an unconditional separation in computational power between shallow quantum and classical circuits. Further results \cite{Shepherd2009,BMS16,bouland_et_al:LIPIcs:2018:8867}, that are based on computational complexity assumptions, show that elementary quantum circuits exist that are likely difficult to simulate on a classical computer. While these results are encouraging, we need to continue researching quantum circuit complexity to point towards meaningful quantum applications that offer a speed up over classical approaches.

A more systematic path towards developing quantum applications for near-term quantum devices, that exhibit a reliable advantage, is based on the complexity theoretic hardness of quantum circuits. In this approach it is the quantum circuit that determines the application, placing the formal complexity result at the beginning of the development.

\subsection{Quantum Machine Learning}
One example of a quantum-classical hybrid algorithm that relies on quantum circuits believed to scale inefficiently for classical methods has been presented in \cite{Havlicek18}. In this work the authors describe and implement two methods of binary classification using supervised training. These classification algorithms are related to standard Support Vector Machines (SVM). The idea in this work, is to implement a non-linear feature map that brings the data to classify into a space in which it can be linearly separated. The key aspect exploited by a quantum processor is that the feature map is implemented as a quantum circuit, mapping the initial data to the high - dimensional quantum state space, so it can be separated by a linear binary classifier data. The use of a quantum feature map has recently also been proposed in \cite{Schuld19}. For this algorithm to provide a quantum advantage, a quantum circuit has to be used that has transition amplitudes that can not be estimated classically to an additive sampling error. The feature map circuit used in~\cite{Havlicek18} can be related to a hardness result derived in \cite{Aaronson18} which guarantees an exponential separation in query complexity to the best classical algorithm. 

In \cite{Havlicek18} two methods are explored to construct a binary classifier based on the hard feature map circuit. In the first method, the feature map circuit is directly followed by a variational circuit. The circuit can be used as a classifier that implements a binary measurement on the quantum feature space. As such this variational algorithm is directly related to a classical SVM. The second method directly exploits the connection to classical SVMs by estimating the Kernel matrix directly on the quantum computer and then using a conventional SVM. The hardware implementation of these two methods showed that even on a modest quantum processor, some sort of error mitigation~\cite{Kandala18} was needed. We have discussed a few error mitigation proposals in Section~\ref{error_mitigation}.

A key observation of this proposal has been that there exist quantum circuits that give rise to feature maps that are hard to evaluate classically, relative to complexity theoretic assumptions. However, to obtain a quantum advantage for a practically relevant machine learning problem such a hard feature map circuit is only a necessary condition. To make this sufficient, more circuits need to be explored that can be tied to complex real world classification problems. 

\subsection{Quantum Chemistry}
A second example uses quantum-classical hybrid algorithms with short-depth quantum circuits for quantum chemistry. The variational quantum eigensolver (VQE)~\cite{peruzzo2014} has been implemented on a number of different quantum hardware platforms~\cite{peruzzo2014,Wang15,omalley2016,Kandala17,Kandala18,Shen17,hempel2018}. Here the central objective is to obtain a good estimate of the ground state energy for a chemistry, or general many-body Hamiltonian. Although, typically fermionic Hamiltonians are considered, these Hamiltonians can be readily mapped to qubit / spin - degrees of freedom by a common procedures ~\cite{bravyi2017tapering}. To obtain ground state energy estimates for a target Hamiltonian, variational trial states are prepared on a quantum computer by using shallow circuits with free parameters that are experimentally adjustable.
The quantum computer is used to estimate the mean energy of the target Hamiltonian by measuring the operators of the Hamiltonian directly on the quantum computer with respect to the trial state. The energy associated with each trial state is then fed to a classical computer, which runs an optimization routine that supplies a new set of parameters. The goal of the optimization procedure it to prepare a new trial state that will tend to lower the energy. This process is then iterated until some convergence condition is met. 

In this approach to near-term quantum algorithms, the utility of the quantum computer lies in the preparation of trial states and the measurement of associated expectation values. Depending on the trial state, these tasks that may be hard on a classical computer, e.g. \cite{schuch2007computational,schwarz2013preparing}. Preparing and measuring trial states on a quantum computer can provide additive error approximations to expectation values. However, as highlighted previously, the algorithm can only yield a quantum advantage if the circuits employed for trial state preparation are difficult to simulate classically. Most VQE implementations to date have focused on small ($<$ 10 qubit) molecular Hamiltonians in quantum chemistry.These implementations have employed circuits that implement ``hardware-efficient" trial states ~\cite{Kandala17,Kandala18} or a unitary coupled cluster (UCC) ansatz~\cite{romero2018}. While the UCC approach offers a structured ansatz that maintains physical symmetries, the hardware-efficient circuits employ interactions that are native to the quantum hardware. Other important considerations for the practical implementation of VQE for quantum chemistry are qubit-efficient fermionic mapping schemes~\cite{bravyi2017tapering}, robustness of classical optimizers to hardware noise, and most importantly, the effect of decoherence~\cite{Temme17,Li17,Kandala18} and the measurement cost for molecular Hamiltonians~\cite{Gokhale2019,izmaylov2019}.
\section{Conclusions} 
This article describes the various challenges and opportunities associated with near-term quantum systems, highlighting the necessary components to bring practical quantum computers closer to reality. A unique interplay between hardware, hardware access, software, benchmarking, applications, and error mitigation techniques is required to develop a quantum system capable of one day executing practical calculations or simulations. 

We have also laid out our software approach, which is heavily user-oriented. We feel strongly that a healthy user base will be a guiding force, helping to shape the technical direction for future quantum devices. We have presented the toolset that is made available (i.e., various access levels, SDKs, Qiskit, etc.), and we observe appreciable demand for such tools. Integrating those tools with stable hardware is a significant effort, but we feel it is worth the challenge, as we are exploring and developing systems that have not yet been built!

It is duly acknowledged, however, that the full potential of near-term quantum systems is presently unknown. While no fundamental roadblocks have yet materialized, it is possible that the application range lags behind some of the more enthusiastic expectations. We are optimistic that near-term quantum systems will be capable of at least shedding new light on unexplored physics lying just out of reach for modern simulation tools, or other derivative applications related to, for example, quantum sensing. There are many areas to study and explore, and by giving the right access and tools to researchers, we hope to accelerate the pace of discovery.

On a broader scale, research in the quantum realm continues to fascinate the minds of many researchers. Active research areas not discussed in this article include fault-tolerant quantum computing~\cite{Terhal15,Campbell17,Gambetta17} (e.g., both theory and experiment for small and large demonstrations of fault-tolerant operations), detailed superconducting qubit hardware considerations~\cite{Gambetta17,Devoret13,Krantz19,Wendin17} (e.g., implementations of qubit gates, coherence times, qubit packaging, qubit measurement, etc.) or other quantum hardware approaches (references within, for example, \cite{Gambetta17}), quantum information~\cite{Nielsen00,wilde_2013} (including its relationship with complexity theory \cite{Watrous2009}), quantum communication~\cite{bennett2014,Gisin07} (e.g., QKD), quantum sensing~\cite{Degen17}, and post-quantum cryptography~\cite{Bernstein17}. Taken together, we feel confident that quantum science as a whole will shape our future technology one day, likely in ways we can't currently foresee.

%\begin{figure}[htbp!]
%		 \centering
%	\includegraphics[width=0.5\textwidth]{figs/CRP}
%		 \caption{Schematic of a superconducting qubit fabricated on top of a Silicon substrate. Surface loss can arise due to contributions from the substrate-air (SA) interface, the substrate-metal (SM) interface and the metal-air (MA) interface.}
%		 \label{fig:fig1}
%\end{figure}

\bibliography{references}

\end{document}